\documentclass[twocolumn,aps,amsmath,amssymb,nopacs,floatfix]{revtex4}
\usepackage{epsfig}
\usepackage{graphicx,bm}
\usepackage{amsmath,amssymb,psfrag}
\newcommand{\bxi}{{\bm{\xi}}}

\begin{document}
\title{Cavity Approach to the Spectral Density of Sparse Symmetric Random Matrices}
\author{Tim Rogers}\affiliation{Department of Mathematics, King's College London, Strand, London WC2R 2LS}
\author{Koujin Takeda}\affiliation{ Department of Computational Intelligence and Systems Science, Tokyo Institute of
Technology, Yokohama 226-8502, Japan}
\affiliation{Department of Mathematics, King's College London, Strand, London WC2R 2LS}
\author{Isaac P\'{e}rez Castillo}\affiliation{Department of Mathematics, King's College London, Strand, London WC2R 2LS}
\author{Reimer K\"uhn}\affiliation{Department of Mathematics, King's College London, Strand, London WC2R 2LS}

\begin{abstract}
The spectral density of various ensembles of  sparse symmetric random matrices is analyzed using the cavity method. We consider two cases: matrices whose associated graphs are locally tree-like, and sparse covariance matrices. We derive a closed set of equations from which the density of eigenvalues can be efficiently calculated. Within this approach, the Wigner semicircle law for  Gaussian matrices and the  Mar\u{c}enko-Pastur law for covariance matrices are recovered easily. Our results are compared with numerical diagonalization, finding excellent agreement.
\end{abstract}

\maketitle

\section{Introduction}
What started as an approximation to the complex Hamiltonian of heavy nuclei has become a very interesting area of research in its own right. Although the statistical properties of random matrices had been tackled before, it was that treatment by Wigner in nuclear physics during 1950's which boosted the research of what is currently known as Random Matrix Theory (RMT) \cite{Mehta1991}. The list of applications of this theory has been expanding ever since, ranging from physics, to computer science and finance. Specifically, applications in physics include nuclear theory, quantum
chaos, localization, theory of complex networks, and more (see, for instance, \cite{Guhr1998} for an extensive review).\\
From a theoretical and practical viewpoint one of the central quantities of interest in RMT is the spectral density of an ensemble of random matrices. While some cases have been completely analyzed during the last decades, many others have not been fully explored. Consider, for instance, the ensemble of  symmetric random matrices whose entries are independently and identically distributed Gaussian variables. Among many of its properties, it is well-known that its spectral density is given by the Wigner semicircle law \cite{Wigner,Mehta1991,edwardsjones}. Another instance is the ensemble of covariance matrices, whose spectral density is given by the Mar\u{c}enko-Pastur law \cite{Marcenko1967}. And the list continues.\\
Interestingly, the  change of introducing sparsity in such ensembles, \textit{i.e.} many entries being zero, complicates the mathematical analysis enormously \cite{dorogovtsev,rodgersbray,biroli,cugliandolo,nagao}. Lacking more powerful mathematical tools, one must rely on approximative schemes to the spectral density, \textit{e.g.} the Effective Medium Approximation (EMA), the Single Defect Approximation (SDA) \cite{biroli,cugliandolo,nagao}.\\
In this work, we tackle the problem of evaluating the spectral density of sparse random matrices by using the cavity method \cite{MPV,mezard}. As we will show, this approach may offer new theoretical and practical advantages: from a theoretical point of view, it offers an alternative, and we believe easier, method to (re)derive the spectral density; practically, the resulting cavity equations can be interpreted as a belief-propagation algorithm on single instances, which can be then easily implemented.  The resulting spectral density is a clear improvement over those obtained by approximative schemes. A complementary study using the replica method, and emphasizing ensemble aspects has appeared elsewhere \cite{Reimer}.\\
This work is organized as follows: in Sec. \ref{section:cavity} we first mention how the spectral density can be recast as a problem of interacting particles on a graph. The subsequent problem is then analyzed by the cavity method in two cases: locally-tree like graphs and sparse covariance matrices. We derive cavity equations for large single instances and check that the dense limit gives the correct results. In Sec. \ref{sect:numerics} we use the cavity equations as an algorithm to calculate the spectral density and compare these results with numerical diagonalization.  The last section is for conclusions.

%%%%%%%%%%%%%%%%%%%%%%%%%%%%%%%%%%%%%%%%%%%%%%%%%%%%%%%%%%%%%%%%%%%%%%%%%%
\section{Cavity approach to the spectral density}
\label{section:cavity}
Consider an ensemble $\mathcal{M}$ of  $N\times N$ symmetric matrices. If we denote  with $\{\lambda^A_{i}\}_{i=1,\ldots,N}$ as the set of eigenvalues of a given matrix $A\in\mathcal{M}$, its spectral density is defined as follows 
\begin{equation}
\varrho_{A}(\lambda)=\frac{1}{N}\sum_{i=1}^N\delta(\lambda - \lambda_i^A).
\label{eq:spectral}
\end{equation}
The spectral density of the ensemble, denoted as $\rho(\lambda)$ results from averaging $\varrho_A(\lambda)$ over  $\mathcal{M}$. \\
As it was shown by Edwards and Jones  \cite{edwardsjones}, the density \eqref{eq:spectral} can be rewritten as
\begin{equation}
\varrho_{A}(\lambda)=-\lim_{\epsilon\rightarrow 0^+}\frac{2}{\pi N}\text{Im}\left[\frac{\partial}{\partial  z }\log\mathcal{Z}_{A}(z)\right]_{ z =\lambda-i\epsilon},
\end{equation}
where
\begin{equation}
 \mathcal{Z}_{A}(z)=\int\left[\prod_{i=1}^N\frac{dx_i}{\sqrt{2\pi}}\right]e^{-\frac{1}{2}\sum_{i,j=1}^N x_i(z I  -A)_{ij}x_j}.
\label{eq:partition}
\end{equation}
In writing the expression \eqref{eq:partition}, we have been careless with the Gaussian integrals, so that as they stand they are not generally convergent; we simply follow the prescription as in \cite{Mezard1999,Brezin2006}, and do not worry unnecessarily about imaginary factors, so that we can introduce a Gibbs-Boltzmann probability distribution of $\bm{x}$, \textit{viz.}
\begin{equation}
\begin{split}
P_A(\bm{x})=\frac{1}{\mathcal{Z}_A(z)}e^{-  \mathcal{H}_A(\bm{x},z)}
\label{eq:Gibbs}
\end{split}
\end{equation}
with 
\begin{equation}
 \mathcal{H}_A(\bm{x},z)=\frac{1}{2}\sum_{(i,j)\in\mathcal{G}_{A}}^Nx_i( z  I-A)_{ij}x_j.
\label{eq:Hamiltonian}
\end{equation}
In this way, the spectral density $\varrho_{A}(\lambda)$ is recast into a statistical mechanics problem of $N$ interacting particles $\bm{x}=(x_1,\ldots, x_N)$ on a graph $\mathcal{G}_A$ with effective Hamiltonian \eqref{eq:Hamiltonian}. By $\mathcal{G}_{A}$ we refer to a weighted graph with $N$ nodes and edges for each interacting pair $(i,j)$ with weight $A_{ij}$, when $A_{ij}\neq 0$. For later use, we  introduce the following notation: the set of neighbors of a node $i$ will be denoted as $\partial i$; for a given subset of nodes $\mathcal{B}$ we define $\bm{x}_{\mathcal{B}}=(x_{\ell_1},\ldots,x_{\ell_{|\mathcal{B}|}})$ with $\ell_1,\ldots,\ell_{|\mathcal{B}|}\in\mathcal{B}$ and with $|\mathcal{B}|$ the number of nodes of $\mathcal{B}$;  $k_i=|\partial i|$ denotes the number of neighbors of node $i$, while $c=\frac{1}{N}\sum_{i=1}^N k_i$ is the average connectivity.\\
Note that within this approach, Eq. \eqref{eq:spectral} for the spectral density $\varrho_A(\lambda)$ can be rewritten as follows:
\begin{equation}
\begin{split}
 \varrho_A(\lambda)&=\lim_{\epsilon\rightarrow 0^+}\frac{1}{\pi N}\sum_{i=1}^N \text{Im}\left[\langle x_i^2\rangle_{z}\right]_{z=\lambda-i\epsilon}\,,
\label{eq:secondmoment}
\end{split}
\end{equation}
where $\langle \cdots\rangle_{z}$ denotes average over  distribution \eqref{eq:Gibbs}.\\
In previous works \cite{edwardsjones,rodgersbray,biroli,cugliandolo,nagao}, the averaged spectral density $\rho(\lambda)$ is dealt with by using the replica approach, or in \cite{Fyodorov1991,Rodgers1990} using Supersymmetric methods. For general sparse matrices, it was shown in \cite{rodgersbray,biroli,cugliandolo,nagao,Fyodorov1991} that the analysis of the resulting equations from  either the replica or the Supersymmetric methods  was a rather daunting task. To push the analysis further, the authors in \cite{cugliandolo,nagao} resorted to a series of approximative schemes, originally introduced in \cite{biroli}. The simpler of such approximations, the EMA, assumes that all nodes are equivalent and play the same role \cite{biroli,cugliandolo,nagao}. This type of approximation works better the larger the average connectivity $c$ of the graph. However, for low and moderate values of $c$ it fails to provide an accurate description of central part and the tails  (see, for instance, \cite{cugliandolo}) and of the presence of weighted Dirac delta peaks in the spectral density \cite{golinelli, bauer}. Other approximations, like the SDA, also fail to give an accurate description of the spectrum.\\
To improve our understanding of spectral properties of sparse matrices, we tackle the problem from a different perspective. Instead of considering the averaged spectral density $\rho(\lambda)$, we note, as shown in Eq. \eqref{eq:secondmoment}, that to calculate $\varrho_A(\lambda)$ we simply need the local marginals  $P_{i}(x_i)$  from the Gibbs-Boltzmann distribution $P_{A}(\bm{x})$. The cavity method offers a way to calculate them \footnote{ This approach have been used in \cite{Ciliberti2004} within the context of Anderson localisation.}. To illustrate this we consider two cases: the ensemble of symmetric locally tree-like sparse matrices, and the ensemble of sparse covariance matrices.
\subsection{Tree-like symmetric matrices}
\begin{figure}
\includegraphics[width=250pt]{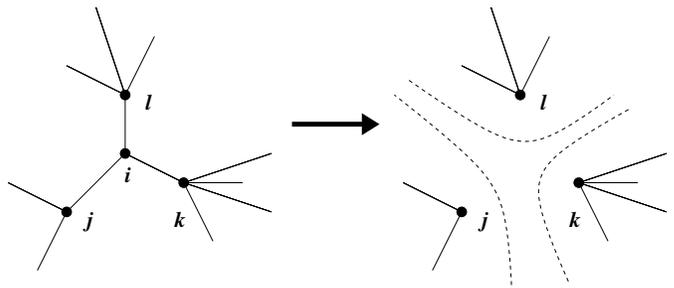}
\caption{Left: Part of a tree-like graph $\mathcal{G}_A$ showing the neighborhood of node $i$. Right: Upon removal of node $i$, on the resulting cavity graph $\mathcal{G}^{(i)}_A$ , the neighboring sites $j$,$k$ and $l$ become uncorrelated.}
\label{fig:one}
\end{figure}

Let us start by analyzing the spectral density of sparse graphs $\mathcal{G}_{A}$ which are tree-like, as the one depicted in Fig. \ref{fig:one}. By tree-like we mean  short loops are rare. 
Due to the tree-like structure we note that for each node $i$, the joint distribution of its neighborhood $P(\bm{x}_{\partial i})$ is correlated mainly through the node $i$. If, instead of the original graph $\mathcal{G}_A$, we consider a system where the node $i$ is removed (see Fig. \ref{fig:one}), on the resulting cavity graph $\mathcal{G}^{(i)}_A$  the joint distribution $P^{(i)}(\bm{x}_{\partial i})$  factorizes, \textit{i.e.}
\begin{equation}
P^{(i)}(\bm{x}_{\partial i})=\prod_{\ell\in\partial i}P^{(i)}_\ell(x_\ell)\,.
\end{equation}
This factorisation, which is exact on trees, is called Bethe approximation. On the cavity graph, the set of cavity distributions  $\{P^{(j)}_i(x_i)\}$ obeys simple recursive equations, \textit{viz.}
\begin{equation}
\begin{split}
 P^{(j)}_i(x_i)&=\frac{e^{-\frac{1}{2} z x_i^2}}{Z_i^{(j)}}\int d \bm{x}_{\partial i\setminus j}e^{x_i\sum_{\ell\in\partial i\setminus j} A_{i\ell}x_\ell} \prod_{\ell\in\partial i\setminus j} P^{(i)}_\ell(x_\ell) \,
\label{eq:cavity}
\end{split}
\end{equation}
for all $i=1,\ldots, N$ and for all $j\in \partial i$. Once the cavity distributions are known, the marginal distributions $P_{i}(x_i)$ of the original system $\mathcal{G}_{A}$ are given by
\begin{equation}
\begin{split}
 P_i(x_i)&=\frac{e^{-\frac{1}{2} z x_i^2}}{Z_i}\int d \bm{x}_{\partial i}e^{x_i\sum_{\ell\in\partial i} A_{i\ell}x_\ell} \prod_{\ell\in\partial i} P^{(i)}_\ell(x_\ell) \,
\label{eq:real}
\end{split}
\end{equation}
for all $i=1,\ldots, N$. While there is {\em in general\/} no a-priori reason to expect cavity distributions to be simple, they {\em are\/} for the present system: the set (\ref{eq:cavity}) of equations is clearly self-consistently solved by Gaussian $P_i^{(j)}$s. Hence, upon assuming the cavity distributions to be Gaussian, namely,
\begin{equation}
 P^{(i)}_\ell(x)=\frac{1}{\sqrt{2\pi\Delta_\ell^{(i)}}}e^{-\frac{1}{2\Delta_\ell ^{(i)}}x^2}
\end{equation}
the set of equations \eqref{eq:cavity}  is transformed into a set of equations for the cavity variances $\Delta_j^{(i)}( z )$, \textit{viz.} 
\begin{equation}
 \Delta_i^{(j)}( z )=\frac{1}{ z -\sum_{\ell\in\partial i\setminus j}A_{i\ell}^2\Delta_\ell^{(i)}( z )}
\label{eq:cavityvar}
\end{equation}
for all $i=1,\ldots, N$ and for all $j\in \partial i$. Similarly, by Eq. \eqref{eq:real} the marginals $P_{i}(x_i)$ are Gaussian with variance $\Delta_i$ related to the cavity variances by
\begin{equation}
 \Delta_i( z )=\frac{1}{ z -\sum_{\ell\in\partial i}A_{i\ell}^2\Delta_\ell^{(i)}( z )}\,.
\label{eq:realvar}
\end{equation}
Eqs. \eqref{eq:cavityvar} and \eqref{eq:realvar} comprise the final result.  For a given graph $\mathcal{G}_A$, one iterates the cavity Eqs. \eqref{eq:cavityvar} until convergence is reached. Once the cavity variances are known, the variances $\Delta_i$ are given by Eqs. \eqref{eq:realvar}, from which the spectral density $\varrho_A(\lambda)$ is obtained by
\begin{equation}
 \varrho_A(\lambda)=\lim_{\epsilon\rightarrow 0^+}\frac{1}{\pi N}\sum_{i=1}^N  \text{Im}\left[\Delta_i( z )\right]_{z=\lambda-i\epsilon}\,.
\end{equation}
It is worth noting that equations equivalent to those in \eqref{eq:cavityvar} can be derived by the method described in \cite{Abou1973} and  can be related to self-returning random walks \cite{malioutov,dorogovtsev}.\\
Notice also that the set of cavity equations must be solved for complex $z=\lambda-i\epsilon$, so that the cavity variances are in general complex, and then  the limit $\epsilon\to 0^{+}$ is performed.  Instead, we  perform this limit explicitly in the cavity equations. To do so, we separate these equations into their real and imaginary parts and then do explicitly the limit $\epsilon\to0^{+}$ by na\"ively assuming that the imaginary part is non-vanishing in such a limit (see discussion below). Denoting $(a_i^{(j)},b_i^{(j)})=[\text{Re}(\Delta_i^{(j)}),\text{Im}(\Delta_i^{(j)})]$, we obtain
\begin{equation}
\begin{split}
a_i^{(j)}&=\frac{\lambda-h^{(j)}_i(\bm{a})}{\left(\lambda- h^{(j)}_i(\bm{a})\right)^2+\left(h^{(j)}_i(\bm{b})\right)^2}\\
b_i^{(j)}&=\frac{h^{(j)}_i(\bm{b})}{\left( \lambda-h^{(j)}_i(\bm{a})\right)^2+\left(h^{(j)}_i(\bm{b})\right)^2}
\label{eq:realIm}
\end{split}
\end{equation}
with
\begin{equation}
\begin{split}
h^{(j)}_i(\bm{v})&=\sum_{\ell\in \partial i\setminus j}A_{i\ell }^2v^{(i)}_\ell\,.
\end{split}
\end{equation}
Our results are exact, as long as the average connectivity $c$ of the graphs considered remains finite in the limit $N \to \infty$.
\subsubsection*{Large $c$ limit: The Wigner Semicircle Law}
To assess our approach, note that from the set of equations \eqref{eq:cavityvar} and \eqref{eq:realvar} we can easily recover the Wigner semicircle law in the large $c$ limit. By this limit we understand that the  $k_i\to c$ and $c\to \infty$, and assume that the graph is already ``infinitely'' large \footnote{Alternatively, one could na\"ively take $c\to N$ and then $N\to\infty$. In the complete, or fully connected, graph ($c=N$) the cavity equations are still valid, but the reason for the decorrelation is statistical rather than topological.}. To do this large $c$ limit, we take the entries of the matrix $A$ to be $A_{ij}=J_{ij}/\sqrt{c}$, with $J_{ij}(=J_{ji})$ a Gaussian variable with zero mean and variance $J^2$. From Eqs. \eqref{eq:cavityvar} and \eqref{eq:realvar}, we note that for large $c$, we have that  $\Delta_i^{(j)}(z)=\Delta_{i}(z)+\mathcal{O}(c^{-1})$ \footnote{This is the usual derivation of TAP equations from the cavity equations. In this case the difference between cavity fields and effective fields does not produce an Onsager reaction term.}.  Upon defining
\begin{equation}
 \Delta=\lim_{c\to\infty}\frac{1}{c}\sum_{\ell\in\partial i}\Delta_\ell,
\end{equation}
we obtain that
\begin{equation}
\lim_{c\to\infty}\sum_{\ell\in\partial i}A_{i\ell}^2\Delta_\ell^{(i)}=\lim_{c\to\infty}\frac{1}{c}\sum_{\ell\in\partial i}J_{i\ell}^2\Delta_\ell^{(i)}=J^2 \Delta\,.
\end{equation}
Thus, in the large $c$ limit Eq. \eqref{eq:realvar} yields
\begin{equation}
 \Delta=\frac{1}{z-J^2\Delta}\,,
\end{equation}
which gives the well-known Wigner semicircle law \cite{Wigner}
\begin{equation}
\begin{split}
\rho(\lambda)&=\frac{1}{2\pi J^2}\sqrt{4J^2-\lambda^2}\,.
\end{split}
\end{equation}
\subsection{Covariance matrices}
\begin{figure}
\includegraphics[width=250pt]{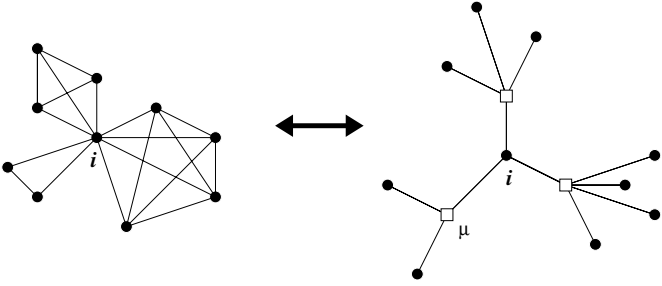}
\caption{Left: Graph $\mathcal{G}_{A}$ for covariance matrices. Right: Bipartite graph $\mathcal{G}_{\bxi}$ for the matrix $\bxi$. For sake of clarity, self-interactions in the graph $\mathcal{G}_{A}$ are not drawn.}
\label{fig:dual}
\end{figure}

Let us consider now matrices  $A$ of the type
\begin{equation}
 A_{ij}=\frac{1}{d}\sum_{\mu=1}^P \xi_{i\mu}\xi_{j\mu}\,,
\label{eq:covariance}
\end{equation}
where $\bxi$ is an $N\times P$ matrix with entries $\xi_{i\mu}$.  To this matrix we can associate a bipartite graph $\mathcal{G}_{\bxi}$ with $N+P$ nodes, divided into two sets indexed by  $i=1,\ldots,N$ and $\mu=1,\ldots,P$ (see Fig. \ref{fig:dual}). A pair of nodes $(i,\mu)$ is connected if $\xi_{i\mu}\neq 0$. We refer to these nodes as $\bm{x}$-nodes and $\bm{m}$-nodes, respectively. We will consider the bipartite graph $\mathcal{G}_{\bxi}$ to be  tree-like, \textit{i.e.} many of the entries $\xi_{i\mu}$ are zero.  We also introduce $d=(1/P)\sum_{\mu=1}^P k_\mu$ with $k_\mu=|\partial \mu|$, \textit{i.e.} the average connectivity of the $\bm{m}-$nodes. Clearly, $c=\alpha d$ with $\alpha=P/N$.
In this case it is more convenient to apply the cavity method on the bipartite graph $\mathcal{G}_{\bxi}$. To do so we write the effective Hamiltonian \eqref{eq:Hamiltonian} as follows
\begin{equation}
\begin{split}
  \mathcal{H}_{A}(\bm{x},z)&= \frac{1}{2}z\sum_{i=1}^N x_i^2 - \frac{1}{2}\sum_{\mu=1}^P m_\mu^2(\bm{x}_{\partial \mu})\,,
\end{split}
\end{equation}
where we have defined the overlaps
\begin{equation}
 m_\mu(\bm{x}_{\partial \mu}) = \frac{1}{\sqrt{d}}\sum_{i\in \partial \mu} \xi_{i\mu}x_i.
\end{equation}
Note that, due to our choice for $\bxi$ and the relation \eqref{eq:covariance} between the matrices $A$ and $\bxi$, the corresponding graph $\mathcal{G}_A$ (see Fig. \ref{fig:dual}) is locally clique-like with self-interactions.\\
We work our cavity equations in the bipartite graph $\mathcal{G}_{\bxi}$. Here, the variables $\bm{x}$ are on the $\bm{x}$-nodes while the variables $\bm{m}=(m_1,\ldots,m_p)$ are on the $\bm{m}$-nodes. Since we have two types of nodes, we apply the cavity method twice: around $\bm{x}$-nodes and around $\bm{m}$-nodes. We define $Q^{(i)}_{\nu}(m_\nu)$ as the cavity distribution of $m_\nu$ in the absence of a node $i$, and $ P^{(\mu)}_i(x_i)$ is the cavity distribution of $x_i$ in the absence of node $\mu$. We find the following set of equations for these cavity distributions
\begin{equation}
\begin{split}
 P^{(\mu)}_i(x_i)&=\frac{e^{-\frac{1}{2}z x_i^2}}{Z_i^{(\mu)}}\int  d\bm{m}_{\partial i\setminus \mu}\,  e^{ \frac{1}{2}\sum_{\nu\in\partial i\setminus \mu} \left(m_\nu+\frac{1}{\sqrt{d}}\xi_{i\nu}x_i\right)^2}\\
&\times\prod_{\nu\in\partial i\setminus \mu} Q^{(i)}_\nu(m_\nu)\,,
\label{eq:cavitycovar1}
\end{split}
\end{equation}
for all $i=1,\ldots, N$ and $\mu\in\partial i$. Also
\begin{equation}
\begin{split}
 Q^{(i)}_{\nu}(m_\nu)&=\frac{1}{Z_\nu^{(i)}}\int d \bm{x}_{\partial \nu\setminus i}\, \delta\left( m_\nu-\frac{1}{\sqrt{d}}\sum_{\ell \in\partial \nu\setminus i}\xi_{\ell\nu}x_\ell \right)\\
&\times \prod_{\ell \in\partial \nu\setminus i} P^{(\nu)}_\ell(x_\ell)\,,
\label{eq:cavitycovar2}
\end{split}
\end{equation}
for all $\nu=1,\ldots, P$ and $i\in\partial \nu$. Obviously, for the marginal distributions $P_i(x_i)$ we obtain
\begin{equation}
\begin{split}
 P_i(x_i)&=\frac{e^{-\frac{1}{2}z x_i^2}}{Z_i}\int  d\bm{m}_{\partial i}\,  e^{ \frac{1}{2}\sum_{\nu\in\partial i} \left(m_\nu+\frac{1}{\sqrt{d}}\xi_{i\nu}x_i\right)^2}\\
 &\times\prod_{\nu\in\partial i} Q^{(i)}_\nu(m_\nu)\,,
\label{eq:realcovar}
\end{split}
\end{equation}
for all $i=1,\ldots, N$. As before, we see from the set of equations \eqref{eq:cavitycovar1} and   \eqref{eq:cavitycovar2} that the Gaussian measure is a fixed point. Thus, by taking $P^{(\mu)}_i(x_i)$ and $Q^{(i)}_\mu(m_\mu)$ to be Gaussian distributions with zero mean and variances $\Delta_i^{(\mu)}$ and $\Gamma_\mu^{(i)}$, respectively, we obtain the following set of equations for the cavity variances
\begin{equation}
\left\{\begin{split}
 \Delta_i^{(\mu)}(z) &= \frac{1}{z - \frac{1}{d}\sum_{\nu\in\partial i\setminus \mu}\xi_{i\nu}^2\frac{1}{1-\Gamma_\nu^{(i)}(z)}}\\
 \Gamma_\nu^{(i)}(z) &= \frac{1}{d}\sum_{\ell \in\partial \nu\setminus i}\xi_{\ell\nu}^2\Delta_\ell^{(\nu)}(z)\,.
\end{split}\right.
\label{eq:Pastur1}
\end{equation}
Similarly, if we denote with $\Delta_i$ the variance of the marginal $P_i(x_i)$ on the original graph $\mathcal{G}_{\bxi}$ we obtain
\begin{equation}
\Delta_i(z) = \frac{1}{z - \frac{1}{d}\sum_{\nu\in\partial i}\xi_{i\nu}^2\frac{1}{1-\Gamma_\nu^{(i)}(z)}}\,.
\label{eq:Pastur2}
\end{equation}
\subsubsection*{Large $c$ limit: The Mar\u{c}enko-Pastur law}
For the sake of simplicity we take the non-zero entries of the matrix $\bxi$ to have values $\pm 1$, so that $\xi_{i\mu}^2=1$ in Eqs. \eqref{eq:Pastur1} and \eqref{eq:Pastur2}. Let us consider  the spectral density in the  large $c$ limit of the bipartite graph $\mathcal{G}_{\bxi}$. By this limit we mean $k_\mu\to d$, $k_i\to c$ and $d,c\to \infty$ while $\alpha$ remains finite. As before, the difference between cavity variances and variances is $\mathcal{O}(c^{-1})$. If we define
\begin{equation}
\begin{split}
\Delta&=\lim_{d\to\infty} \frac{1}{d}\sum_{\ell \in\partial \nu}\Delta_\ell=\lim_{d\to\infty} \frac{1}{d}\sum_{\ell =1}^d\Delta_\ell\,,
\end{split}
\end{equation}
 from Eq. \eqref{eq:Pastur2} we obtain
\begin{equation}
\begin{split}
 \Delta = \frac{1}{z - \alpha\frac{1}{1-\Delta}}\,.
\end{split}
\end{equation}
Upon solving this equation for $\textrm{Im}(\Delta)$ we obtain the Mar\u{c}enko-Pastur law \cite{Marcenko1967} of dense covariance matrices
\begin{equation}
\begin{split}
 \rho( \lambda ) &= \frac{1}{2\pi\lambda}\sqrt{-\lambda^2 + 2\lambda(\alpha+1)+(\alpha-1)^2}\\
&\hspace{10mm} + C_0(1-\alpha) \delta( \lambda )\,.
\end{split}
\end{equation}
with $C_0=1$ for $\alpha\leq 1$ and $C_0=0$ for $\alpha>1$.  A slightly different expression is found by Nakanishi and Takayama \cite{takayama}, where the difference comes from not considering the diagonal terms. This could also be implemented fairly straightforwardly to obtain the spectral density as in \cite{takayama}.
\section{Numerical Results and Comparison}
\label{sect:numerics}
For general sparse matrices, we solve the cavity equations numerically and compare the results with exact numerical diagonalization. We consider again the two cases of locally tree-like and sparse covariance matrices.
\subsection{Tree-like symmetric matrices}
\begin{figure}
\includegraphics[width=230pt]{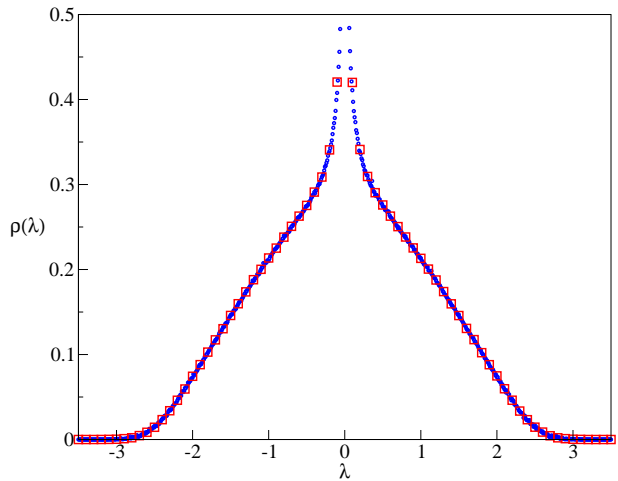}
\caption{\label{fig:G} Spectral density of Poissonian graphs with Gaussian edge weights and average connectivity $c=4$. Red Square markers are for the results of numerical diagonalization with $N=1000$, averaged over $1000$ samples. Blue circles are results of the cavity approach  with $N=1000$ and average over $1000$ samples.}
\end{figure}

To test our cavity equations, we choose Poissonian graphs $\mathcal{G}_{A}$ where each entry $A_{ij}$ of the  $N\times N$ matrix $A$ is drawn independently from
\begin{equation}
 P(A_{ij})= \frac{c}{N}\pi(A_{ij})+ \left(1-\frac{c}{N}\right)\delta(A_{ij})
\end{equation}
with $c$ the average connectivity, and $\pi(x)$ is the distribution of non-zero edge weights. For the distribution of weights we study two cases: bimodal distribution, \textit{i.e.}
\begin{equation}
 \pi(A_{ij})=\frac{1}{2}\delta(A_{ij}-1)+\frac{1}{2}\delta(A_{ij}+1)
\end{equation}
and Gaussian distribution with zero mean and variance $1/c$.\\
For the purpose of fairly comparing later  with exact numerical diagonalization, we have analyzed the cavity equations for rather small matrices. However, we have checked that the convergence of these equations is generally fairly fast and we are able to evaluate the spectral density of very large matrices in reasonable time. In both, the bimodal and the Gaussian cases,  we generated matrices with $N=1000$. For each matrix we run our cavity Eqs. \eqref{eq:cavityvar} until convergence is reached and then obtain the spectral density from Eqs. \eqref{eq:realvar}. The result is averaged over 1000 samples. For such sizes we have also calculated the spectral density by exact numerical diagonalization and averaged over 1000 samples.\\
\begin{figure}
\includegraphics[width=230pt]{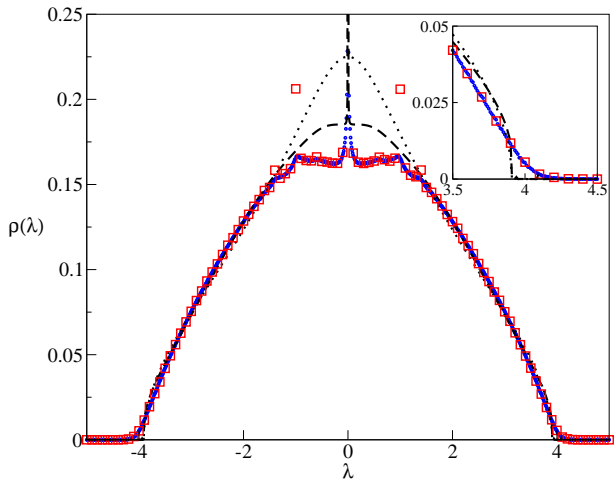}
\caption{\label{fig:B3} Spectral density of Poissonian graphs with bimodal edge weights and average connectivity $c=3$.  Red Square markers are the results of numerical diagonalization with $N=1000$, averaged over $1000$ samples. Blue circles are the results of the cavity approach  with $N=1000$, averaged over $1000$ samples. The dashed line corresponds to the SDA and dotted line is the EMA. The inset shows the tail of the spectral density.}
\end{figure}
The numerical results from the cavity approach and exact  numerical diagonalization for the bimodal case is plotted in Fig. \ref{fig:B3}  for average connectivity $c=3$. We have also compared our results with the spectral density obtained by EMA and SDA (see \cite{cugliandolo, biroli, dorogovtsev} for details about the approximations). As we can see, our results are a clear improvement over the EMA and SDA results, as they are in excellent agreement with numerical diagonalization. Even the tail of the spectrum, usually not obtained with these approximation, (see inset of Fig. \ref{fig:B3}) is well reproduced by our approach.\\
It is well known that the spectrum of these type of ensembles contains a dense collection of Dirac delta peaks \cite{golinelli, bauer}, which are not fully captured by the previous approximations. Without a prior analysis, one wonders how the cavity equations can be used to obtain such contributions. A practical way out is to reconsider the limit $\epsilon\to 0$, by leaving a small value of $\epsilon$ in the cavity equations, which  implies approximating Dirac deltas by Lorentzian peaks.
\begin{figure}
\includegraphics[width=230pt]{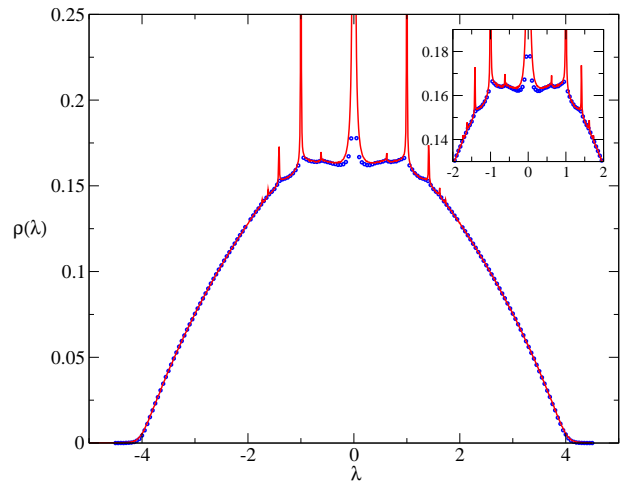}
\caption{\label{fig:B3_v2} Comparison of cavity equations for $\epsilon=0$ (blue circles) and $\epsilon=0.005$ (Continuous red line), for Poissonian graphs with bimodal edge weights and average connectivity $c=3$ ($N=1000$ and average over $1000$ samples). The inset shows the  Dirac delta structure in the central region.}
\end{figure}
 In Fig. \ref{fig:B3_v2} we have rerun the set of eqs \eqref{eq:cavityvar} and \eqref{eq:realvar} with a small value of $\epsilon$. The Dirac delta contributions, whose exact positions within the spectral density is discussed in \cite{golinelli}, are now clearly visible. A more detailed study on this issue within the context of localisation can be found in \cite{Reimer}.\\
In Fig. \ref{fig:G} we plot the results of both numerical diagonalization and the cavity method when $\pi(x)$ is a Gaussian distribution with zero mean and variance $1/c$. Once again, our results are in excellent agreement with the numerical simulations.\\

\subsection{Covariance matrices}
\begin{figure}
\includegraphics[width=230pt]{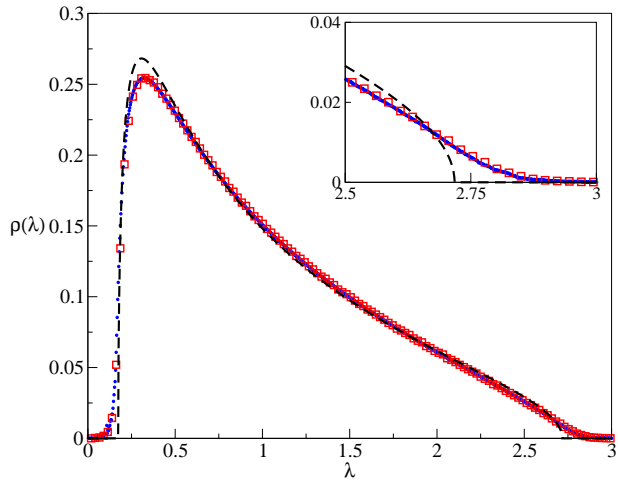}
\caption{Spectral density of  covariance matrices with $N=4000$, $d=12$, $\alpha=0.3$. Average over 1000 samples. Red Square markers are for the results of numerical diagonalization. Blue circles are results of the cavity approach. The dashed line corresponds to the SEMA.}
\label{fig:H}
\end{figure}

We have also analyzed numerically in the case of  sparse covariance matrices. Here the entries $\xi_{i\mu}$ of the $N\times P$ matrix $\bxi$, are drawn according to the distribution
\begin{equation}
  P(\xi_{i\mu})=\frac{d}{N}\pi(\xi_{i\mu})+\left(1-\frac{d}{N}\right)\delta(\xi_{i\mu})\, ,
\end{equation}
where $\pi(\xi_{i\mu})$ is a bimodal distribution
\begin{equation}
\pi(\xi_{i\mu})=\frac{1}{2}\delta(\xi_{i\mu}+1)+\frac{1}{2}\delta(\xi_{i\mu}-1)
\end{equation}
In Fig. \ref{fig:H},  we compare the results of direct diagonalization, the cavity method, and the symmetric effective medium approximation (SEMA), introduced in \cite{nagao}; here we make the same choice of parameters. The inset figure shows detail of the tail region of the plot, where the difference between the SEMA and the other results can be clearly seen.
\section{Conclusions}
In this work, we have re-examined the spectral density of ensembles of sparse random symmetric matrices. By following Edwards and Jones \cite{edwardsjones}, we have mapped the problem into an interacting system of particles on a sparse graph, which was then analyzed by the cavity approach. Within this framework, we have derived cavity equations on single instances and used them to calculate the spectral densities of sparse symmetric matrices. Our results are in good agreement with numerical diagonalization and are a clear improvement to previous works based on approximative schemes. We have also shown that, to account for the Dirac delta contribution to the spectrum, one may approximate Dirac delta peaks by Lorentzians, by leaving a small value for $\epsilon$ \cite{Reimer}.\\
It is well known that cavity and replica methods are equivalent (see for instance \cite{mezard} for diluted spin glasses), so one may wonder in which aspects our work differs from the ones presented in \cite{rodgersbray,biroli,cugliandolo,nagao}. Generally, for interacting diluted systems with continuous dynamical variables one expects an infinite number of cavity fields to parametrize the cavity distributions. The authors in \cite{rodgersbray,biroli,cugliandolo,nagao} decided to tackle such a daunting task by resorting to approximations. In this work, we simply realize that, for the problem at hand, the cavity distributions are Gaussian, so that the problem can be solved exactly by self-consitently determining the variances of these distributions.\\
In future studies we expect to extend the method presented here to the analysis of more general aspects of random matrices.

\acknowledgments
KT thanks the hospitality of the Disordered Systems Group, at the department of Mathematics, King's College. He is supported by Grand-in-aid from MEXT/JSPS, Japan (No.18079006) and Program for Promoting Internationalization of University Education, MEXT, Japan (Support for Learning Overseas Advanced Practices in Research). The authors thank M M\'ezard and Y Kabashima for discussions and G Parisi for discussions and for pointing out earlier work on the subject. IPC also thanks ACC Coolen for his work during the initial stages of this Guzai project.

\bibliography{CavityMethod}

\end{document}